\documentclass[12pt]{iopart}
\usepackage{graphicx}

\begin{document}


\title{Kolmogorov-Sinai entropy in field line diffusion by anisotropic magnetic turbulence}


\author{Alexander V. Milovanov}
\address{Associazione Euratom-ENEA sulla Fusione, Centro Ricerche Frascati, Via E. Fermi 45, C.P. 65, I-00044 Frascati, Rome, Italy.
Also at: Department of Space Plasma Physics, Space Research Institute, Russian Academy of Sciences, Profsoyuznaya 84/32, 117997 Moscow, Russia}

\author{Rehab Bitane}
\address{Laboratoire Cassiop\'ee, UNSA, CNRS, Observatoire de la Cote d'Azur, 
BP 4229, 06304 Nice Cedex 4, France}

\author{Gaetano Zimbardo}
\address{Dipartimento di Fisica, Universit\'a degli Studi della Calabria, Ponte P. Bucci, Cubo 31C, I-87036 Arcavacata di Rende, Italy}



\begin{abstract}
The Kolmogorov-Sinai (KS) entropy in turbulent diffusion of magnetic field lines is analyzed on the basis of a numerical simulation model and theoretical investigations. In the parameter range of strongly anisotropic magnetic turbulence the KS entropy is shown to deviate considerably from the earlier predicted scaling relations [Rev. Mod. Phys. {\bf 64}, 961 (1992)]. In particular, a slowing down logarithmic behavior versus the so-called Kubo number $R\gg 1$ ($R = (\delta B / B_0) (\xi_\| / \xi_\bot)$, where $\delta B / B_0$ is the ratio of the rms magnetic fluctuation field to the magnetic field strength, and $\xi_\bot$ and $\xi_\|$ are the correlation lengths in respective dimensions) is found instead of a power-law dependence. These discrepancies are explained from general principles of Hamiltonian dynamics. We discuss the implication of Hamiltonian properties in governing the paradigmatic ``percolation" transport, characterized by $R\rightarrow\infty$, associating it with the concept of pseudochaos (random non-chaotic dynamics with zero Lyapunov exponents). Applications of this study pertain to both fusion and astrophysical plasma and by mathematical analogy to problems outside the plasma physics.\\
\\
This research article is dedicated to the memory of Professor George M. Zaslavsky
\end{abstract}


\maketitle

\section{Introduction}

The presence of low-frequency, long-wavelength fluctuations in hot magnetized plasmas is often found to alter the plasma confinement properties leading to an anomalously high heat and energy transfer across magnetic field lines as compared to purely collisional values, a phenomenon known as anomalous or turbulent transport. While the basic deteriorating effect of the fluctuations on the plasma confinement is relatively well understood, a detailed, microscopic picture of anomalous transport turns out to be far than simple. Situations exist in which collisional and fluctuation induced properties operate essentially on the same footing, as for instance some regimes of electron outward fluxes in solar coronal loops as discussed by Galloway {\it et al.} \cite{Galloway}. Well before that, in the fusion field, essentially the same mechanism has been proposed for electron heat transport due to parallel thermal conductivity in a tokamak with destroyed magnetic surfaces \cite{Rosenbluth}. Early investigations of this problem are due to the pioneering work of Rosenbluth {\it et al.} \cite{Taylor}. In all these models the fluctuations are such as to introduce some form of braiding of magnetic field lines \cite{Stix}, with ``time" associated with the axial (toroidal) dependence. In the braided magnetic fields, the collisional transverse diffusion is naturally enhanced as a consequence of the parallel motion providing a continuous mapping of the particle's perpendicular random walk. In realistic tokamaks, in view of the many different processes actually producing the transport (e.g., Refs. \cite{Horton,Dendy,White,Zonca95,Zonca,Carreras}, just to mention some), effects caused by magnetic braiding can, however, represent only a fraction of the overall transport of energy and heat and be significant within only a limited range of parameters. Nevertheless, addressing the magnetic braiding phenomenon in the appropriate setting \cite{Rosenbluth,Taylor} defines a problem of fundamental and practical importance. Not only does it relate to a fascinating set of nonlinear dynamics issues \cite{Sagdeev}, but also a reliable extrapolation of braiding induced transport losses to reactor-relevant plasmas is not at hand.    

In order to accurately assess the collisional transport in the braided magnetic field, a few parameters are needed. These parameters are: (i) parallel ($\xi_\|$) and perpendicular ($\xi_\bot$) fluctuation correlation lengths; (ii) collisional parallel and perpendicular particle diffusivities: these quantities are actually known as functions of plasma density and temperature in both axial and toroidal geometries \cite{Jeffrey}; (iii) fluctuation induced magnetic field line diffusion coefficient ($D_m$); and (iv) the rate of exponential separation of initially close magnetic field lines, $h$, a parameter often referred to as the Kolmogorov-Sinai (KS) entropy \cite{Sagdeev,ZaslavskyUFN,Rechester79,Bickerton97}. 

Most previous theoretical and numerical studies have focused on addressing the scaling laws for the diffusion coefficient $D_m$ as a function of the fluctuation strength or, more conveniently, the Kubo number $R = (\delta B / B_0) (\xi_\| / \xi_\bot)$, where $\delta B / B_0$ is the ratio of the rms magnetic fluctuation field to the background magnetic field. In general, the results of those investigations can be summarized by a power law dependence $D_m \propto R^\gamma$ (Refs. \cite{Isi,Misguich,Vlad,Zim00}), where  the scaling is quasilinear-like ($\gamma = 2$) for $R\ll 1$ and percolation-like ($\gamma = 7/10$) for $R\gg 1$ (as distinct from traditionally believed, Bohm-like scaling with $\gamma = 1$ \cite{Bohm,Taylor71}). More recently, the percolation scaling with $\gamma = 7/10$ was questioned by Milovanov \cite{PRE01,PRE09}, who argued that $\gamma = 2/3$ instead. Support to this value can be found in the numerical simulation results of Ref. \cite{Zim01,PScripta}. Milovanov has also addressed \cite{PRE09} the theoretical foundations of the percolation scaling by relating it to a description in terms of fractional derivative equations and Hamiltonian {\it pseudochaos} (random non-chaotic dynamics with zero Lyapunov exponents) \cite{Report}. This connection to pseudochaotic properties of percolation will play a central role in the theoretical investigations below.   

Actually less progress has been made with respect to addressing the $h$ value. Here, too, scaling relations have been proposed \cite{Isi,Horton2}: $h\propto R^2$ for $R\ll 1$ and $h\propto R^{1/2}\ln R$ for $R\gg 1$. Even so, their direct numerical investigation has been performed only recently \cite{Zimbardo}. A net result of this investigation follows: While the quadratic behavior of $h$ in the parameter range of small Kubo numbers $R\ll 1$ could readily be reproduced, a strong deviation from the predicted square root-like scaling $h\propto R^{1/2}\ln R$ was found for $R\gg 1$. Indeed, the observed behavior was markedly slower (Fig.~4 in Ref. \cite{Zimbardo}), possibly slower than a power law, with a tendency to saturation for the largest achievable $R$ (of order 10$^2$). This study is aimed to elaborate those investigations and to comprehend the observed discrepancies.

The rest of the paper is organized as follows. A numerical study of the problem is presented in Sec. 2, where we detail the numerical techniques and also demonstrate the phenomenon of saturation of the KS entropy for the largest achievable $R$. These results are theoretically analyzed in Sec. 3 on the basis of a Hamiltonian approach. In particular we discuss the implication of fractal geometric properties in the entropy growth slowing down. We summarize our conclusions in Sec. 4.  

\section{Numerical study}

We begin with the standard equations for magnetic field lines, with ${\bf r}$ a three-dimensional (3D) position vector, written as 
\begin{equation}
d{\bf r} / {ds} = \delta {\bf B}({\bf r}) / |{\bf B}({\bf r})|, \label{1} 
\end{equation} 
where $s$ is the natural parameter, such that $\int ds$ is the length along a field line, and ${\bf B}({\bf r})$ is the magnetic field. The latter is set as a sum of a constant homogeneous background field ${\bf B}_0 = B_0 \hat{\bf z}$ in a straight-cylinder geometry (here $\hat{\bf z}$ is a unit vector in the axial direction) and a static magnetic perturbation, defined by its Fourier series expansion as 
\begin{equation}
\delta {\bf B}({\bf r}) = \sum_{\sigma = 1,2}\sum_{{\bf k}} \delta B _{\bf k} {\bf e}^{(\sigma)}_{\bf k} \exp i[{\bf k\cdot r}+\phi_{\bf k}^{(\sigma)}], \label{2} 
\end{equation} 
where ${\bf e}^{(\sigma)}_{\bf k}$ are two polarization unit vectors (these corresponding to the shear Alfv\'en and compressional Alfv\'en waves), $\phi_{\bf k}^{(\sigma)}$ are randomly chosen phases, and $\delta B_{\bf k}$ are the spectral amplitudes. Basically, this is the same
realization of magnetic fluctuations as in Refs. \cite{Zim00,Zimbardo}.
It is assumed that $\delta {B}_{-\bf k} = \delta {B}_{\bf k}$ and $\phi_{\bf -k}^{(\sigma)} = \phi_{\bf k}^{(\sigma)}$, so that the field $\delta {\bf B}({\bf r})$ be real. Finally, the fluctuation spectrum is taken in the power law form to be, with $\Theta = \Theta (\xi_{\bot}, \xi_{\|})$ a normalization constant, 
\begin{equation}
\delta B_{\bf k}={{\Theta (\xi_{\bot}, \xi_{\|})}\over(k_{\bot}^2\xi_{\bot}^2+k_{\|}^2\xi_{\|}^2)^{\alpha /4+1/2}}. \label{3} 
\end{equation}
Such power law spectra of varying anisotropy have been found in many plasmas, both in space and in the laboratory \cite{Burlaga86,JGR96,JASTP,Borovsky,Carreras2,Zaslav,UFN}. In the simulation $\alpha = 5/3$ reminding of the Kolmogorov spectrum of fully developed turbulence. It is worth to remark that, while the exact entropy value in the simulation may be somewhat dependent on $\alpha$, we expect that the basic physics conclusions regarding the functional behavior of $h$ will remain essentially the same. This expectation relies on the results of theoretical analysis below, which will lead to (i) generic functional form for the entropy in Hamiltonian dynamics, and (ii) comprehension of the features of nonlinear saturation of the $h$ function in association with the natural dependency of resonant and percolation properties on the Kubo number.    

Note that the numerical model setup in Eqs.~(\ref{1})$-$(\ref{3}) is essentially 3D. The values of the correlation lengths $\xi_{\|}, \, \xi_\bot$ determine the shape of the constant intensity
ellipsoids in $\bf k$-space. 
Special cases of this correspond to an isotropic turbulence ($\xi_\bot = \xi_{\|}$); a quasi-2D turbulence ($\xi_{\|} \gg \xi_\bot$); and the slab turbulence ($\xi_{\|} \ll \xi_\bot$). For the present analysis, $\xi_{\|} > \xi_\bot$, but other cases have been considered in Refs. \cite{Zim00,Zimbardo}. The wave numbers are taken from $N_{\rm min}^2 \le n_x^2+n_y^2+n_z^2 \le N_{\rm max}^2$, with the wave vectors defined as ${\bf k} = ({2\pi}/{N_{\rm min}}) \left({n_x}/{\xi_{\bot}},{n_y}/{\xi_{\bot}},{n_z}/{\xi_{\|}}\right)$. To match with the natural physics and computer limitations, the power law dependence in Eq.~(\ref{3}) is truncated on both sides implying finite cutoffs at $k_{\bot \rm min} = 2\pi / \xi_{\bot}$ and  $k_{\bot \rm max} = (2\pi / \xi_{\bot}) (N_{\rm max}/N_{\rm min})$ in the perpendicular plane, and at $k_{\|\rm min} = 2\pi / \xi_{\|}$ and $k_{{\|}\rm max} = (2\pi / \xi_{\|}) (N_{\rm max}/N_{\rm min})$ in the axial direction. It is convenient to use $2\pi / k_{\bot\max}$ as the unit length in the perpendicular plane. Hence the perpendicular correlation length can be expressed as $\xi_\bot = N_{\rm max}/N_{\rm min}$. To match with the expected properties of heat transport in coronal loops and pertinent model parameters of Ref. \cite{Zimbardo}, here we set $N_{\rm min}\simeq 2.23$ and $N_{\rm max} = 16$, which generates some 8200 independent wave modes in Eq.~(\ref{2}). We take notice of the fact that the KS entropy may depend on the spectral extension in Eq.~(\ref{3}) since smaller wavelengths favour faster separation of the field lines \cite{Zim84,Zim95,Ruffolo04}; however, for the runs presented here, we keep the spectral extension fixed (i.e., $N_{\rm max} / N_{\rm min}$ is not varied). The sizes (in respective directions) of the simulation box are set to $L_{x,y} = N_{\rm min} \xi_{\bot} = N_{\rm max}$ and $L_z = N_{\rm min} \xi_{\|} = ( \xi_{\|} / \xi_\bot)N_{\rm max}$, so that at least two correlation lengths have to be traveled before the same magnetic configuration is reproduced. This effectively eliminates the periodicity effects \cite{Pommois98}. 

\begin{figure}
\includegraphics[width=0.70\textwidth]{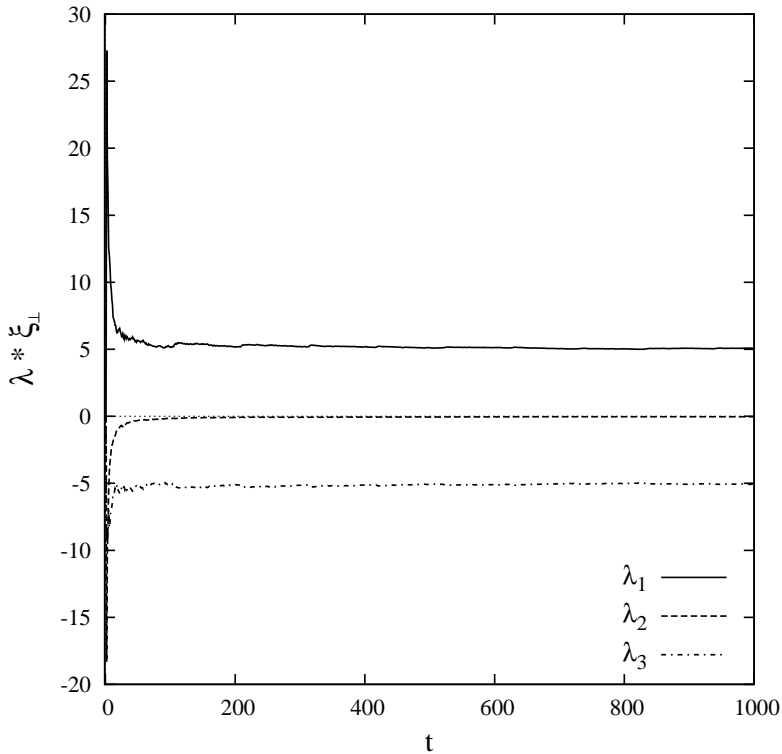} 
\caption{The three Lyapunov exponents computed as a function of time for $R = 40$.
After an initial transient, the exponents settle to the asymptotic values.}
\end{figure}

In the basic theory \cite{Sagdeev,Benettin76}, the KS entropy is obtained as the sum of the positive Lyapunov exponents for the unstable motions. The Lyapunov exponents, in their turn, can be calculated numerically by implementing the algorithm of Wolf {\em et al} \cite{Wolf85}, in which they are obtained from the Gram-Schmidt orthonormalization. 
In the numerical model, $\delta B / B_0$ is varied from 0.01 to 1, and $\xi_{\|}/\xi_{\bot}$ from 1 to 100; this choice of parameters has earlier been discussed for astrophysical applications in Refs. \cite{Zim01,Ruffolo04,Matthaeus03}. With these settings, the Kubo number $R$ can be varied from 0.01 to 100. These values extrapolate from the quasilinear to the percolation regime as discussed in Refs. \cite{Zimbardo,Zim00}. Also, it was shown in Ref. \cite{Zimbardo} that the
value of $h$ depends only on $R$, and not separately on $\delta B/B_0$ or $\xi_{\|} / \xi_\bot$.  
In this work, focusing mainly on the percolation regime, we have restricted our investigations to the parameter range $R > 1$. Overall, from 40 to 100 different initial conditions for the field line Eq.~(\ref{1}) have been used to compute the Lyapunov exponents numerically. The numerical accuracy has been checked by asking that the sum of the three Lyapunov exponents be zero,
as implied by $\nabla \cdot {\bf B} =0$. After an initial transient, this was soon achieved with
an accuracy of the order of $10^{-6}$. We also checked that the numerical algorithm has actually reached the asymptotic regime: Figure 1 shows the three Lyapunov exponents for $R=40$ as a function of the integration time. It can be seen that only one Lyapunov exponent is substantially larger than zero, and that the sum of the three exponents equals zero. Also, some small fluctuations are seen in the long time regime: this allows to estimate the statistical error of
the Kolmogorov entropy, which turns out to be less than 0.1 $\xi_\bot^{-1}$.
The typical results of this investigation obtained for $\xi_{\|} / \xi_\bot =100$ and $1 < R < 100$ are shown in Figure 2. 

There are two main conclusions to be drawn from the analysis. First, for not too large the $R$ values, $1 < \sim R < \sim 30$, the $h$ function is well approximated by a logarithmic dependence $h \simeq p \ln R + q$, with $p$ and $q$ the numerical fitting parameters. This is shown in detail in the inset of Figure 2, where the fit is made only for the points up to $R=40$. Second, with the increasing $R$ the entropy logarithmic growth slows down showing signatures of saturation, consistently with our first results in Ref. \cite{Zimbardo}. This is clearly seen in Figure 2,
as fitting all the points with the logarithmic dependence shows a discrepancy which goes beyond
the numerical error bars. 
When larger $R$ are achieved in the simulation, the slowing down effect is more pronounced. 
These behaviors are in marked contrast with the prediction of Ref. \cite{Isi} that $h\propto R^{1/2}\ln R$ for $R\gg 1$.   

\section{Theoretical model}

\subsection{Hamiltonian formulation}

In the parameter range of practical importance, the rms magnetic fluctuation is supposed not to exceed the background magnetic field $B_0$. To simplify the analysis, we shall assume that $\delta B / B_0 \ll 1$. Being somewhat restrictive, this condition, however, still retains the essential physics of the model. Since we are interested in the regime when $R\gg 1$, we assume that $\xi_\| / \xi_\bot \gg 1$ with a large margin. Thus, it is required that the turbulence be strongly anisotropic and its wave vector spectrum be actually squeezed to the perpendicular plane. With this gross yet relevant simplification we can cast the field-line Eqs.~(\ref{1}) in the Hamiltonian form
\begin{equation}
\frac{dx}{dz} = \frac{\partial H(x,y,z)}{\partial y}, \ \ \ \frac{dy}{dz} = -\frac{\partial H(x,y,z)}{\partial x} \label{4} 
\end{equation} 
by simplifying the magnetic field model to ${\bf B} ({\bf r}) = B_0\hat{\bf z} + \delta {\bf B}_\bot ({\bf r}_\bot,z)$, where $\delta {\bf B}_\bot ({\bf r}_\bot, z) = \nabla A_\| ({\bf r}_\bot, z)\times\hat{\bf z}$ is a static, transverse magnetic perturbation, $A_\| ({\bf r}_\bot, z)$ is the axial component of the vector potential, $\nabla = \partial / \partial {\bf r}_\bot$, and ${\bf r}_\bot = (x,y)$ is the position vector in the perpendicular plane. Above, ``time" is associated with the $z$ coordinate, canonical momentum corresponds to the $y$ coordinate, and
$x$ is the coordiante of a one and a half degrees of freedom system. 
With these settings the Hamiltonian is expressible in terms of the $A_\|$ function as $H({\bf r}_\bot,z) = A_\| ({\bf r}_\bot, z) / B_0$. Hereafter $B_0 = 1$ for mathematical simplicity. The flow velocity is given by ${\bf v}_\bot = \nabla H ({\bf r}_\bot,z)\times\hat{\bf z} = \delta {\bf B}_\bot / B_0$. If one introduces the Kubo number as $R\propto v_\bot/\omega \xi_\bot$, one finds $R\propto (\delta B / B_0) (\xi_\| / \xi_\bot)$, as expected, with $v_\bot \simeq\delta B / B_0 \ll 1$ a characteristic value of $|{\bf v}_\bot|$, and $\omega \propto 1/\xi_\|$ the ``frequency" of turbulence variation due to the $z$ dependence.  

\begin{figure}
\includegraphics[width=0.90\textwidth]{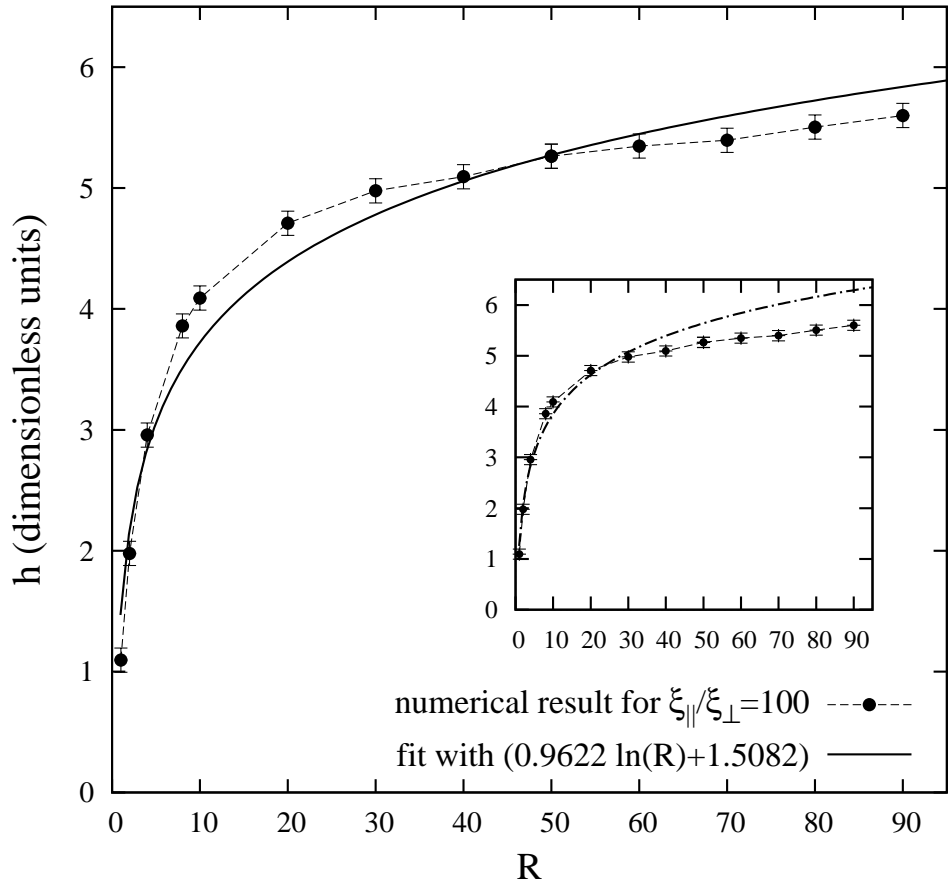} 
\caption{The Kolmogorov-Sinai entropy as a function of the Kubo number in the parameter range $1< R<  90$, with the anisotropy level set to $\xi_{\|} / \xi_\bot =100$. Dots with error bars: the numerically computed $h$ values. Full line: a best fit by a logarithmic function (fit parameters in the legend). The inset shows the same data points, but the logarithmic fit (dash-dotted line) is done for the points up to $R=40$. Note the entropy slowing down for $R> \sim 40$.}
\end{figure}

\subsection{Periodic orbits and separatrices}

To match with the numerical simulation model in Eq.~(\ref{2}), the vector potential is assumed to be given by a sum of plane waves, with the spectral components being odd functions of ${\bf k}_\bot$. At a given time $z$, consistently with the sign symmetry of the wave expansion, the separatrices of the $A_\|$ field are given by $A_\| ({\bf r}_\bot, z={\rm const}) = 0$, while the closed equipotential contours are defined for $A_\| \neq 0$ as 
\begin{equation}
y = y (x, A_\|={\rm const}, z={\rm const}). \label{5} 
\end{equation} 
As usual, the area embraced by a closed contour is obtained as $I = \oint y dx$ (this being the action of the Hamiltonian system). When the equipotentials in Eq.~(\ref{5}) are closed for all $z$, one speaks about a magnetic surface \cite{Taylor}. Magnetic surfaces are labeled by respective $A_\|$ values, while $I = I(A_\|)$ defines the integrals of motion. A field line wrapping around a magnetic surface experiences a periodic motion in the $(x,y)$ plane with the frequency $\Omega (I) = dH (I) / dI = |\nabla A_\| ({\bf r}_\bot, z)| /|\nabla I ({\bf r}_\bot, z)| \simeq v_\bot / \ell$, where $\ell \simeq |\nabla I ({\bf r}_\bot, z)|$ is the perimeter of the equipotential contour. In writing the first term in the sequence, use has been made of the standard definition of the
action-angle variables $(I,w)$ and of the equation for $dw/dt = \Omega (I)$ in Hamiltonian dynamics \cite{Mechanics}. Assuming circular equipotentials one immediately gets $I ({r}_\bot, z)\propto {r}_\bot^2$, $|\nabla I ({r}_\bot, z)| \propto {r}_\bot$, and $I\propto \ell ^2$. 
Here, ${r}_\bot$ represents the typical ``radius" of a closed contour, in the same sense as the
``diameter" of a dust grain. 
In the case of strongly shaped equipotential contours, however, the latter scaling relations may not hold true. Of particular interest are the equipotentials characterized by self-similar geometry in a broad range of spatial scales best described as fractals \cite{Feder}. The significance of such equipotentials in low-frequency turbulence is addressed shortly. For fractal equipotential contours the simple scaling $I\propto \ell ^2$ is generalized to $I\propto \ell ^{2/d_h}$ (the so-called ``area-perimeter relation" \cite{LeMehaute}) where $d_h$ is the fractal dimension of the equipotential line and we have omitted redundant dimensional constants for simplicity. The fractal dimension $d_h$ is defined by the scaling $\ell \propto {r}_\bot^{d_h}$ (instead of $\ell \propto r_\bot$ in non-fractal geometry). The $d_h$ values are generally fractional larger than 1. Eliminating $\ell$ by means of the area-perimeter relation one obtains $\Omega (I) \propto v_\bot / I^{d_h / 2}$.

\subsection{Resonances and the percolation property}

In the basic theory of Hamiltonian chaos it is well established that the existence of resonances has an important impact onto the dynamics \cite{Sagdeev,Zaslavsky,ZaslavskyUFN}. In the model studied here, resonances occur between the wrapping motion of magnetic field lines around the magnetic surfaces and the magnetic field variation due to the axial dependence of the fluctuation field. In the presence of local instability the matching axial and wrapping frequencies or the matching harmonics of these favor departure from the exact periodic, energy conserving motion thus giving rise to transport phenomena in phase space. In the limit of very low frequencies, the conditions for a resonance \cite{Sagdeev} require the corresponding excursion periods to diverge. Mathematically, this can be satisfied for fractal equipotential contours \cite{LeMehaute}, characterized by diverging lengths due to the structure on many scales.

Next, because of the sign symmetry of the fluctuation field in Eq.~(\ref{2}), the zero-set $A_\| ({\bf r}_\bot, z={\rm const}) = 0$ contains a percolating equipotential line, which stretches the entire system \cite{Isi,PRE00}. The percolating line is the channel through which the turbulent diffusion penetrates to the large scales. Analyses of Ref. \cite{PRE00} have shown that the equipotential contours in the model turbulence field~(\ref{2}) lying in close proximity to the percolating line $A_\|\rightarrow 0$ are indeed characterized by the fractal geometry in the limit
when the perpendicular correlation length 
 $\xi_\bot \rightarrow \infty$, and their fractal dimension was found to be $d_h \approx 4/3$, in agreement with the results of Ref. \cite{Aharony}. This fractal geometry is a generic property of fields with a broad power-law energy density distribution conventionally associated with ``turbulence" \cite{UFN}. Thus, when the fluctuations are very slow, we expect the resonance conditions to be naturally satisfied in vicinity of the percolating line, provided that the number of waves in Eq.~(\ref{2}) is sufficiently large. In order words, fractality emerging from the percolation property of $A_\| ({\bf r}_\bot, z={\rm const}) = 0$, together with the wide spectral extension in Eq.~(\ref{3}), guarantees the {\it existence} of the resonant orbits for $\omega\rightarrow 0$ thus supporting the turbulent diffusion in the limit $R\rightarrow\infty$. 

\subsection{Nonlinearity parameter}

A net result from the above reasoning is that the resonance conditions in the limit of very low frequencies (large Kubo numbers) are satisfied for fractal equipotential contours, which, as is already mentioned above, are found in close proximity to the percolating line. The resonance conditions alone do not, however, define the microscopic behaviors. To this end, two important conditions need to be specified. One is the size of the nonlinearity parameter \cite{Sagdeev,Zaslavsky}
\begin{equation}
\Lambda = \frac{I}{\omega}\left|\frac{d\Omega (I)}{dI}\right|. \label{7} 
\end{equation} 
In order for phase space diffusion to actually occur, this parameter must be large, i.e., $\Lambda \gg 1$. In this range, as was first observed by Chirikov, the resonances will overlap \cite{ZaslavskyUFN} and the dynamics can, after memory for the initial conditions has been lost, be described statistically in terms of a Fokker-Planck equation \cite{Zaslavsky}. Topologically, the loss of memory is due to the mixing of trajectories in phase space. The mixing time is obtained from $z_c \simeq 1 / 2\omega\ln\Lambda$, while the number of overlapping resonances is ordered as $\sqrt{\Lambda} \gg 1$. 

\subsection{Chaotic and pseudochaotic properties}

Next the relationship between mixing and randomness must be addressed, which can be nontrivial in some cases. Indeed, because of the many overlapping resonances, the dynamics will, in fact, be random after a certain transient time, which one would estimate as, say, two-three characteristic times it takes to pass from one resonance to the other. When the phase space accessible for random motions is wide enough (ideally, comparable with the entire phase space of the system), then the transition to random dynamics occurs on time scales of the order of $z_c$. This is a typical situation in chaotic dynamics. It should be understood, however, that this typical situation does not always take place. Situations exist, in which the available phase space is just too narrow to permit good mixing properties. For instance, in low-frequency turbulence, random motions as dictated by pertinent resonance conditions can be squeezed to a subset of phase space with fractal geometry \cite{PRE09}. It is this a situation that occurs in critical percolation \cite{Isi}. As a consequence, the dynamics only partially acquire the chaotic property in that they show random behavior, while the mixing turns out to be very poor (sub-exponential). This anomalously slow mixing may lead to ``strange" phenomena in phase space, for instance to stickiness of phase space trajectories \cite{Report}, just to mention one. The implication is that the Lyapunov exponents may vanish despite of the inherently random character of the motion. This type of behavior has come to be known as {\it pseudochaotic} \cite{Report,JMPB,PD2004}. In pseudochaos one finds, then, that the transition to random dynamics occurs on time scales much shorter than $z_c$. A simple model realization of this is found in random walks on percolation systems \cite{PRE09}. Anticipating our final conclusion below, here we mention that the observed properties of saturation of the KS entropy are explained in terms of the Hamiltonian pseudochaos.  

\subsection{Order of limits}

The other relevant condition that one needs to identify the microscopic pattern of behavior brings together the two building blocks of the model: fractality and randomness. Focusing on the fractal properties we somehow expect the fractal range to be wide enough, ideally $\xi_\bot \rightarrow\infty$. On the other hand, the slowness of the fluctuations implies that $R\propto v_\bot / \omega\xi_\bot\rightarrow\infty$ for $\omega\rightarrow 0$. For the purpose of formal orderings, because the two limiting procedures may not be commutative, we require $\xi_\bot\rightarrow\infty$ diverge faster than $R\rightarrow\infty$. That is, given a finite fluctuation frequency, we set $\xi_\bot \gg v_\bot / \omega\xi_\bot$. The implication is that the fluctuations should not be too slow for the actual (finite) size of the system, otherwise the dynamics retain a deterministic character. Equivalently, $\xi_\bot^2 / v_\bot \gg 1/\omega \propto \xi_\|$, showing that the diffusion time to the distance $\xi_\bot$ must be large compared to the period of the field. Under these conditions, which are naturally satisfied for the small enough $v_\bot \simeq\delta B / B_0 \ll 1$, the dynamics will be inherently random with $\Lambda \gg 1$ even for very slow fluctuation frequencies.  

\subsection{Scaling argument}

Let us now rewrite Eq.~(\ref{7}) in a more insightful form. Substituting $\Omega (I) \propto v_\bot / I^{d_h / 2}$ and performing the trivial differentiation over $I$ leads to $\Lambda (\ell) \propto v_\bot / \omega\ell$, where the area-perimeter relation has been applied. For the longest equipotentials present ($r_\bot \sim \xi_\bot$), the perimeter scales as $\ell \propto \xi_\bot^{d_h}$, making it possible to evaluate $\Lambda \propto v_\bot / \omega\xi_\bot^{d_h}$. Remembering that the Kubo number $R\propto v_\bot / \omega\xi_\bot$ it is found that
\begin{equation}
R/\Lambda \propto \xi_\bot^{d_h - 1}. \label{8} 
\end{equation} 
This is a very important relation because it reveals that the Kubo number $R$ can be very large if the geometry is fractal ($\xi_\bot \gg 1$, $d_h > 1$). When $d_h \rightarrow 1$, fractal properties are lost, while the $R$ and $\Lambda$ parameters turn out to be of the same order.  

\subsection{Kolmogorov-Sinai entropy $-$ functional form}

We are now in position to obtain the KS entropy for the system under consideration. In fact, the entropy, $h$, can be evaluated as the inverse mixing time, $z_c$ (called by some \cite{Galloway,Bickerton97} the Kolmogorov length $L_K$to emphasize its geometric meaning), yielding $h\simeq 1/z_c \simeq 2\omega\ln\Lambda$ (Ref. \cite{Sagdeev} for the full discussion). Eliminating $\Lambda$ by means of Eq.~(\ref{8}) it is found that
\begin{equation}
h\simeq 2\omega\ln R - 2\omega(d_h - 1)\ln\xi_\bot - 2\omega\ln C, \label{9} 
\end{equation} 
where the last term in the sequence arises from the suppressed dimensional constants in the above scaling relations. Given the anisotropy level, $\xi_\| / \xi_\bot$, Eq.~(\ref{9}) defines the KS entropy in function of the fluctuation strength or the Kubo number, $R$. It is understood that, by changing $R$, one also changes the contour or set of the contours on which the resonances occur. Fractal geometries of the resonant contours may also be different for the different $R$ values. In this sense one should think of the $d_h$ parameter in Eq.~(\ref{9}) as being function of $R$.   

\subsection{Chaotic case}

The next step is to observe that the resonance conditions cannot simultaneously be satisfied for the fractal and non-fractal equipotential contours because of the immense difference in their lengths. The actual regime is determined by the degree of anisotropy of the turbulence, with stronger anisotropy favoring the implication of fractal properties. Consider first the non-fractal case; that is, assume the anisotropy be kind of moderate. Setting $d_h = 1$ in Eq.~(\ref{8}), one finds $R\simeq\Lambda$. Thus, in the non-fractal regime, a large $R\gg 1$ implies a correspondingly large $\Lambda\gg 1$, and, hence, the behavior is chaotic \cite{Zaslavsky}. This conclusion finds support in the numerical simulation results of Ref. \cite{Zim00}, where Poincar\'e sections of field-line motion have been analyzed (Figs.~2 and~3 in their work). For $d_h = 1$, the KS entropy in Eq.~(\ref{9}) reduces to $h\simeq 2\omega\ln R - 2\omega\ln C$. This logarithmic dependence on $R$ is consistent with the numerical fit in the inset of Figure 2 for not too large the $R$ values.

\subsection{Nonlinear saturation and pseudochaos}

With the increasing anisotropy (i.e., Kubo number) the resonant motions gradually shift to fractal equipotential contours characterized by $d_h > 1$. This causes the $h$ value to deviate from a purely logarithmic growth, since the second term in Eq.~(\ref{9}) is now non-zero. Thus, the behavior is actually slower than a logarithm, as can be seen by close inspection of Figure 2. Clearly, the entropy growth slowing down occurs as a consequence of fractality influencing the resonance conditions. 

Furthermore, there is an upper bound on the entropy growth, and that is where the ``saturation" \cite{Zimbardo} of the entropy comes into play. When the turbulence becomes extremely anisotropic, i.e., $\xi_\| / \xi_\bot\rightarrow\infty$, the resonance conditions are satisfied in a very small neighborhood of the separatrix, $A_\| \rightarrow 0$. In this small vicinity, the resonance structure is actually very dense. If one defines the characteristic frequency, $\omega$, one finds the width of the resonance layer to be $\delta\omega\propto\sqrt{\omega}$ (as compared to the distance between the resonances $\Delta\omega\propto\omega$). See Ref. \cite{Sagdeev}. Note that $\delta\omega\gg\Delta\omega$ for $\omega\rightarrow 0$, and that the width of the resonance layer shrinks to zero (as $\sqrt{\omega}$), while the density of the resonances in the layer, defined as inverse distance $\Delta\omega$, diverges as $1/\omega$. 

In a sense, $\omega$ being close to zero implies that all the resonances have overlapped in a very narrow stochastic layer enveloping the separatrix. Let the number of these resonances be ${\mathcal {N}}_{\max}\gg 1$. Since the number of overlapping resonances is given by $\sqrt{\Lambda}$, it is noted that the $\Lambda$ value has reached its maximum; that is, $\Lambda_{\max}\simeq({\mathcal {N}}_{\max})^2$. Hence, for a given (finite) $\xi_\|$, the KS entropy saturates at $h_{\max}\simeq 2\omega\ln\Lambda_{\max} \propto (4/\xi_\|)\ln{\mathcal {N}}_{\max}$. When $\xi_\| \rightarrow \infty$, the resonance spectrum collapses, while the entropy $h_{\max} \rightarrow 0$. 

Under the conditions specified above, the dynamics will inherently be random despite of the vanishing $h_{\max}$. This means, precisely, that the asymptotic ($R\rightarrow\infty$, $h_{\max} \rightarrow 0$) dynamics are {\it pseudochaotic}, in the sense of Refs. \cite{Report,PRE09,JMPB,PD2004}. To this end, our final conclusion is: The deviation from a purely logarithmic behavior and the ultimate nonlinear saturation of the KS entropy in the percolation limit are explained by the natural tendency of the considered system to evolve into a dynamical state characterized by the pseudochaotic property.

\subsection{Connection to fractional derivative equations}

The dynamics, being pseudochaotic in the percolation limit, needs to be addressed. Because of the formally diverging mixing time $z_c\simeq 1/h_{\max}\rightarrow\infty$ the Fokker-Planck equation in the form of Ref. \cite{Zaslavsky} can be written only asymptotically for $z\gg z_c$. On time scales shorter than $z_c$ this conventional form of the Fokker-Planck equation is invalidated. There is a growing belief that, in many ways, pseudochaotic systems are described by space-time self-similarity and by kinetic equations written in terms of generalized (fractional) derivatives \cite{Report}. We note in passing that a fractional extension of the Fokker-Planck equation has, for Hamiltonian systems, been proposed in early works in Refs. \cite{PhysicaD,Chaos}.

More so, because there is no unambiguous definition of the notion of fractional derivative, one important issue is to define in which sense the fractional operator is understood in the problem under consideration. In this spirit, Milovanov \cite{PRE09} has suggested that the percolation transport in Hamiltonian systems is described by a fractional time Fokker-Planck equation with the fractional derivative in the Caputo sense (as opposed to the alternative definition in the Riemann-Liouville sense) \cite{Podlubny,Oldham}. This suggestion conforms with the results of Refs. \cite{Castillo1,Castillo2}, where one finds a discussion of subtleties of definition of fractional derivatives and their relevance in plasma physics applications.      

\subsection{Summary of the discussion}

Our main conclusion from the above analysis is: The entropy growth slowing down is due to the concentration of the random properties on a substrate of phase space with fractal geometry in response to the increasing anisotropy of the turbulence. The asymptotic behavior in the limit $R\rightarrow\infty$ and the paradigmatic concept of ``percolation" transport \cite{Isi} conform with the notion of Hamiltonian pseudochaos (random non-chaotic dynamics with zero Lyapunov exponents) \cite{Report}.

\section{Conclusions}

In the present work, we have addressed the Kolmogorov-Sinai (KS) entropy, $h$, in turbulent diffusion of magnetic field lines in the parameter range of large Kubo numbers, $R\gg 1$. For not too large the $R$ values (in practice, $1< R< 30$), we found numerically that the $h$ function is well approximated by a logarithmic dependence $h \simeq p \ln R + q$ (here, $p$ and $q$ are numerical fitting parameters). Most importantly, with the increasing $R$, the entropy logarithmic growth slows down showing signatures of saturation. These behaviors are in marked contrast with the prediction of Ref. \cite{Isi} that $h\propto R^{1/2}\ln R$ for $R\gg 1$. 

To comprehend this apparent discrepancy, we have proposed a model of percolation transport of magnetic field lines, basing our investigations on general principles of Hamiltonian dynamics. We found that the logarithmic counterpart of the entropy is naturally explained by the Hamiltonian chaotic dynamics of magnetic field lines in the sense of Refs. \cite{Sagdeev,Zaslavsky,ZaslavskyUFN}. The deviation from the purely logarithmic behavior and the slowing down of the entropy is explained by the concentration of the chaotic properties on a subset of phase space with fractal geometry. This fractal geometry is a generic property of fields with a broad power-law energy density distribution conventionally associated with the notion of ``turbulence" \cite{UFN}. Finally, the asymptotic saturation of the entropy is explained by the collapse of the resonance spectrum of the system in the limit $\omega\rightarrow 0$ (i.e., $\xi_\|\rightarrow\infty$). 

At this point, a subtlety has occurred involving the order of limits (i.e., the fractal limit $\xi_\bot \rightarrow\infty$ versus $R\rightarrow\infty$). The convention adopted in this work is that $\xi_\bot \rightarrow\infty$ diverges faster than $R\rightarrow\infty$. Under this last condition, the dynamics retain the random character despite of the vanishing KS entropy. This behavior bears signatures enabling to associate it with the notion of {\it pseudochaos} (random non-chaotic dynamics with zero Lyapunov exponents) \cite{Report,JMPB,PD2004}. 

The final result of this investigation is that the paradigmatic ``percolation" transport in the limit $R\rightarrow\infty$ involves two essential ingredients: randomness and fractality. Random motions are found to pursue in the limit $\omega\rightarrow 0$ supported by the fractal geometry of the substrate on which the turbulent transport concentrates. It is the interplay of these two ingredients that explains the observed properties of behavior of the KS entropy. These analyses shed new light on the origin of percolation transport by associating it with a pseudochaotic (rather than simply chaotic) behavior. A formal treatment of the problem in terms of random walk models and fractional kinetic equations is given in Ref. \cite{PRE09}.

\ack
This work was funded in part by the Italian INAF, the Italian Space Agency under the contract ASI n. I/015/07/0 ``Esplorazione del Sistema Solare," and by the INTAS research project 06-1000017-8943. A.V.M. acknowledges the warm hospitality of the Department of Physics
of the University of Calabria, where most of this work was carried out.


\vskip 5pt


\begin{thebibliography}{}

\bibitem{Galloway}
R. K. Galloway,  P. Helander, and A. L. MacKinnon,  Astrophys. J. {\bf 646}, 615 (2006).

\bibitem{Rosenbluth}
A. B. Rechester and M. N. Rosenbluth, Phys. Rev. Lett. {\bf 40}, 38 (1978). 

\bibitem{Taylor}
M. N. Rosenbluth, R. Z. Sagdeev, J. B. Taylor, and G. M. Zaslavsky, Nucl. Fusion {\bf 6}, 297 (1966).

\bibitem{Stix}
T. H. Stix, Phys. Rev. Lett. {\bf 30}, 833 (1973).

\bibitem{Horton}
W. Horton, Rev. Mod. Phys. {\bf 71}, 735 (1999).

\bibitem{Dendy}
G. Manfredi and R. O. Dendy, Phys. Plasmas {\bf 4}, 628 (1997).



\bibitem{White}
R. White, L. Chen, and F. Zonca, Phys. Plasmas {\bf 12}, 057304 (2005).

\bibitem{Zonca95}
F. Zonca, S. Briguglio, L. Chen, G. Fogaccia, and G. Vlad, Nucl. Fusion {\bf 45}, 477 (2005).

\bibitem{Zonca}
F. Zonca, S. Briguglio, L. Chen, G. Fogaccia, T. S. Hahm, A. V. Milovanov, and G. Vlad, Plasma Phys. Controlled Fusion {\bf 48}, B15 (2006).

\bibitem{Carreras}
B. A. Carreras, V. E. Lynch, P. Ph. van Milligen, and R. S\'anchez, Phys. Plasmas {\bf 13}, 062301 (2006).

\bibitem{Sagdeev}
G. M. Zaslavsky and R. Z. Sagdeev, {\it Introduction to the Nonlinear Physics. From Pendulum to Turbulence and Chaos} (Nauka, Moscow, 1988).

\bibitem{Jeffrey}
J. Freidberg, {\it Plasma Physics and Fusion Energy} (Cambridge Univ. Press, Cambridge, 2007).

\bibitem{ZaslavskyUFN}
G. M. Zaslavsky and B. V. Chirikov, Phys. Usp. {\bf 14}, 549 (1972). 

\bibitem{Rechester79}
A. B. Rechester, M. N. Rosenbluth, and R. B. White, Phys. Rev. Lett. {\bf 42}, 1247 (1979). 

\bibitem{Bickerton97}
R. J. Bickerton, Plasma Phys. Controlled Fusion {\bf 39}, 339 (1997).

\bibitem{Isi}
M. B. Isichenko, Rev. Mod. Phys. {\bf 64}, 961 (1992).

\bibitem{Misguich}
J.-D. Reuss and J. H. Misguich, Phys. Rev. E {\bf 54}, 1857 (1996).

\bibitem{Vlad}
J.-D. Reuss, M. Vlad, and J. H. Misguich, Phys. Lett. A {\bf 241}, 94 (1998).

\bibitem{Zim00}
G. Zimbardo, P. Veltri, and P. Pommois, Phys. Rev. E {\bf 61}, 1940 (2000).

\bibitem{Bohm}
T. H. Dupree, Phys. Fluids {\bf 10}, 1049 (1967).

\bibitem{Taylor71}
J. B. Taylor and B. McNamara, Phys. Fluids {\bf 14}, 1492 (1971). 

\bibitem{PRE01}
A. V. Milovanov, Phys. Rev. E {\bf 63}, 047301 (2001).

\bibitem{PRE09}
A. V. Milovanov, Phys. Rev. E (in the press, accepted paper in Plasma Physics). e-print arXiv:0903.3534.

\bibitem{Zim01}
P. Pommois, P. Veltri, and G. Zimbardo, Phys. Rev. E {\bf 63}, 066405 (2001).

\bibitem{PScripta}
F. Chiaravalloti, A. V. Milovanov, and G. Zimbardo, Phys. Scr. {\bf T122}, 79 (2006).

\bibitem{Report}
G. M. Zaslavsky, Phys. Rep. {\bf 371},  461 (2002).

\bibitem{Horton2}
M. B. Isichenko, W. Horton,  D. E Kim, E. G. Heo, and D.-I. Choi, Phys. Fluids B {\bf 4}, 3973 (1992).

\bibitem{Zimbardo}
G. Zimbardo, R. Bitane, P. Pommois, and P. Veltri, Plasma Phys. Controlled Fusion {\bf 51}, 015005 (2009).

\bibitem{Burlaga86}
L. F. Burlaga and L. W. Klein, J. Geophys. Res. {\bf 91}, 347 (1986). 

\bibitem{JGR96}
A. V. Milovanov, L. M. Zelenyi, and G. Zimbardo, J. Geophys. Res. {\bf 101}, 19 903 (1996). 

\bibitem{JASTP}
A. V. Milovanov, L. M. Zelenyi, P. Veltri, G. Zimbardo, and A. L. Taktakashvili, J. Atmos. Solar Terr. Phys. {\bf 63}, 705 (2001).

\bibitem{Borovsky}
J. M. Weygand, M. G. Kivelson, K. K. Khurana, H. K. Schwarzl, S. M. Thompson, R. L. Mc.Pherron, A. Balogh, L. M. Kistler, M. L. Goldstein, J. Borovsky, and D. A. Roberts, J. Geophys. Res. {\bf 110}, A01205 (2005).

\bibitem{Carreras2}
B. A. Carreras, B. van Milligen, C. Hidalgo, R. Balbin, E. Sanchez, I. Garcia-Cortes, M. A. Pedrosa, J. Bleuel, and M. Endler, Phys. Rev. Lett. {\bf 83}, 3653 (1999). 

\bibitem{Zaslav}
G. M. Zaslavsky, M. Edelman, H. Weitzner, B. Carreras, G. McKee, R. Bravenec, and R. Fonck, Phys. Plasmas {\bf 7}, 3691 (2000). 

\bibitem{UFN}
L. M. Zelenyi and A. V. Milovanov, Phys. Usp. {\bf 47}, 749 (2004). 

\bibitem{Pommois98}
P. Pommois, G. Zimbardo,  and P. Veltri, Phys. Plasmas {\bf 5}, {1288} (1998).

\bibitem{Zim84}
G. Zimbardo,  P. Veltri, and F. Malara, J. Plasma Phys. {\bf 32}, 141 (1984).

\bibitem{Zim95}
G. Zimbardo, P. Veltri,  G. Basile, and S. Principato, Phys. Plasmas {\bf 2}, 2653 (1995).

\bibitem{Ruffolo04}
D. Ruffolo, W. H. Matthaeus,  and P. Chuychai, Astrophys. J. {\bf 614}, {420} (2004).

\bibitem{Benettin76}
G. Benettin, L. Galgani, and J. M. Strelcyn, Phys. Rev. A {\bf 14}, 2338 (1976).

\bibitem{Wolf85}
A. Wolf, J.B. Swift, H. Swinney, and A. Vastano, Physica D {\bf 16}, 285 (1985).

\bibitem{Matthaeus03}
W. H. Matthaeus, G. Qin, J. W. Bieber, and G. P. Zank, Astrophys. J. Lett. {\bf 590}, {L53} (2003).

\bibitem{Mechanics}
L. D. Landau and E. M. Lifshitz, {\it Course of Theoretical Physics. Vol. 1. Mechanics} (Pergamon Press, Oxford, 1969). 

\bibitem{Feder}
J. Feder, {\it Fractals} (Plenum, New York, 1988). 

\bibitem{LeMehaute}
A. Le Mehaute, {\it Fractal Geometries: Theory and Applications} (CRC Press, Boca Raton, FL, 1991).

\bibitem{Zaslavsky}
G. M. Zaslavsky, {\it Statistical Irreversibility in Nonlinear Systems} (Nauka, Moscow, 1970).

\bibitem{PRE00}
A. V. Milovanov and G. Zimbardo, Phys. Rev. E {\bf 62}, 250 (2000).

\bibitem{Aharony}
T. Grossman and A. Aharony, J. Phys. A {\bf 19}, L745 (1986).

\bibitem{JMPB}
O. Lyubomudrov, M. Edelman, and G. M. Zaslavsky, Intl. J. Modern Phys. B {\bf 17}, 4149 (2003).  

\bibitem{PD2004}
G. M. Zaslavsky and M. A. Edelman, Physica D {\bf 193}, 128 (2004).

\bibitem{Chaos}
G. M. Zaslavsky, Chaos {\bf 4}, 25 (1994).

\bibitem{PhysicaD}
G. M. Zaslavsky, Physica D {\bf 76}, 110 (1994).

\bibitem{Podlubny}
I. Podlubny, {\it Fractional Differential Equations} (Academic Press, San Diego, 1999). 

\bibitem{Oldham}
K. B. Oldham and J. Spanier, {\it The Fractional Calculus} (Academic Press, San Diego, Calif. 1974).

\bibitem{Castillo1}
D. del-Castillo-Negrete, B. A. Carreras, and V. E. Lynch, Phys. Plasmas {\bf 11}, 3854 (2004).

\bibitem{Castillo2}
D. del-Castillo-Negrete, Phys. Plasmas {\bf 13}, 082308 (2006).


\end{thebibliography}
\end{document}